\documentclass[preprint,showpacs,amsmath,amssymb,aps,floats,floatfix,nofootinbib]{revtex4-1}
\usepackage{amsmath,amsfonts,amssymb}
\usepackage{graphicx}
\usepackage{color}
\usepackage[dvipsnames,svgnames,x11names]{xcolor}
\usepackage{slashed} 
\usepackage{url}
\usepackage{subfigure}
\usepackage{multirow} 

\def\cm{\mathrm{cm}} 
\def\km{\mathrm{km}} 
\def\kpc{\mathrm{kpc}} 
\def\sec{\mathrm{s}} 
\def\TeV{\mathrm{TeV}} 
\def\GeV{\mathrm{GeV}} 

\makeatother

\allowdisplaybreaks 

\begin{document}

\title{ Implications for dark matter annihilation from the AMS-02 $\bar{p}/p$ ratio}
\author{Su-Jie Lin, Xiao-Jun Bi, Peng-Fei Yin, Zhao-Huan Yu}
\affiliation{Key Laboratory of Particle Astrophysics,
Institute of High Energy Physics, Chinese Academy of Sciences,
Beijing 100049, China}

\begin{abstract}
The AMS-02 collaboration has just released the cosmic antiproton to proton ratio $\bar{p}/p$ with a high precision up to $\sim 450$~GeV. In this work, we calculate the secondary antiprotons generated by cosmic ray interactions with the interstellar medium taking into account the uncertainties from the cosmic ray propagation. The $\bar{p}/p$ ratio predicted by these processes shows some tension with the AMS-02 data in some regions of propagation parameters, but the excess is not significant. We then try to derive upper bounds on the dark matter annihilation cross section from the $\bar{p}/p$ data or signal regions favored by the data. It is shown that the constraint derived by the
AMS-02 data is similar to that from Fermi-LAT observations of dwarf galaxies. The signal region for dark matter is usually required $m_\chi \sim O(10)$ TeV and $\left<\sigma v\right>\sim\mathcal{O}(10^{-23})~\cm^3~\sec^{-1}$.
\end{abstract}


\maketitle


\section{Introduction}

Antimatter search in cosmic rays (CRs) are especially important for astrophysics and dark matter signal study. In recent years, there were great progresses in the measurements of CR antimatter particles. In particular, the Alpha Magnetic Spectrometer (AMS-02) launched in 2011 has provided unprecedentedly precise measurement of CRs. Based on the results from AMS-02 and previous experiments, such as PAMELA and Fermi-LAT, the properties of CR antimatter particles have been extensively studied in a quantitative way in the literature.

Recently, the AMS-02 collaboration has reported their measurement of the $\bar{p}/p$ ratio up to $\sim 450$~GeV with a high precision~\cite{AMSantip}.
The previous PAMELA $\bar{p}/p$ ratio data~\cite{Adriani:2008zq,Adriani:2010rc} seemed to be compatible with the conventional CR prediction in the energy range of $\mathcal{O}(1)-\mathcal{O}(10)$~GeV, and can be used to set limits on the DM annihilation cross section or lifetime~\cite{Donato:2008jk,Cirelli:2013hv,Bringmann:2014lpa,Cirelli:2014lwa,Jin:2014ica,Hooper:2014ysa,Evoli:2011id,Boudaud:2014qra}.
However, it is very interesting to note that the AMS-02 $\bar{p}/p$ ratio becomes almost flat from $10$~GeV to several hundred GeV. This behavior is not expected from the CR secondary production since the measured B/C ratio decreases with energy as a power-law spectrum without any structure. Therefore it is necessary to perform a careful study on the secondary production of antiprotons taking all the uncertainties into account.

The major challenge to interpret the AMS-02 data is how to give an accurate prediction of the secondary antiprotons. Although the AMS-02 collaboration has provided precise measurements of the proton and helium spectra, there are still large uncertainties from CR propagation processes. The complicated CR propagation in the Galaxy may involve diffusion, energy loss, convection, and reacceleration effects. The propagation parameters describing these effects can be determined by fitting the secondary-to-primary ratios, such as B/C and (Sc+Ti+V)/Fe, and unstable-to-stable ratios of secondary particles, such as $^{10}\mathrm{Be}/^9\mathrm{Be}$ and $^{26}\mathrm{Al}/^{27}\mathrm{Al}$~\cite{Strong:1998pw,cosmicraybook,DiBernardo:2009ku,Maurin:2001sj}.
However, current measurements of these ratios are still not sufficient to distinguish different propagation models. Another important uncertainty comes from the solar modulation, which significantly affects low energy CR spectra, but it is not easy to quantify the effects.

In this work, we give first a careful analysis of the secondary antiproton production. The GALPROP package is used to solve the CR propagation equation~\cite{Strong:1998pw,Moskalenko:1997gh}. We adopt two kinds of propagation models, namely the diffusion-reacceleration (DR) model and diffusion-convection (DC) model. Propagation parameters are determined by fitting the available B/C data with a Markov Chain Monte Carlo (MCMC, \cite{Lewis:2002ah}) algorithm~\cite{Liu:2009sq,Liu:2011re,Yuan:2013eja,Lin:2014vja}, which is powerful for surveying high dimensional parameter space. Secondary antiprotons from CR interactions with the ISM are calculated and the uncertainties from propagation effects are obtained. Taking all the uncertainties into account, the secondary antiprotons seem marginally consistent with AMS-02 results, but show clearly different slope with energy.

Then we study the implications of the AMS-02 $p/\bar{p}$ data on DM annihilation signals. Since the AMS-02 data do not unambiguously indicate an excess, we derive upper limits on the DM annihilation rate and signal regions where the background plus DM signals give a good fit to the AMS-02 data. In order to derive upper limits or signals regions, we carry out a global fitting to the AMS-02 data with the MCMC method. Based on the AMS-02 $\bar{p}/p$ data, upper limits on the DM annihilation rate and favored DM annihilation parameters have also been given in Refs.~\cite{Giesen:2015ufa,Jin:2015sqa,Ibe:2015tma,Hamaguchi:2015wga}.

This paper is organized as follows. In Sec.~II, we describe the CR propagation processes. In Sec.~III, we first determine the propagation parameters from the AMS-02 B/C ratio data
and then calculate the secondary antiproton flux. In Sec.~IV, we study implication on DM annihilation by the AMS-02 $\bar{p}/p$ ratio data. Finally, we give the discussions and conclusions in Sec.~V.

\section{Propagation of Galactic cosmic rays}

After accelerated in sources, Galactic CRs are injected and diffuse in the interstellar space and suffer from several propagation effects before they arrive at the Earth. A conventional assumption is that CRs propagate in a cylindrical halo with a half height $z_h$, beyond which CRs escape freely.
The propagation equation can be expressed as~\cite{Strong:2007nh}
\begin{eqnarray}
\frac{\partial \psi}{\partial t} &=& Q(\mathbf{x}, p) + \nabla \cdot ( D_{xx}\nabla\psi - \mathbf{V}_{c}\psi )
+ \frac{\partial}{\partial p}\left[p^2D_{pp}\frac{\partial}{\partial p}\frac{\psi}{p^2}\right]
\nonumber\\
&& - \frac{\partial}{\partial p}\left[ \dot{p}\psi - \frac{p}{3}(\nabla\cdot\mathbf{V}_c)\psi \right]
- \frac{\psi}{\tau_f} - \frac{\psi}{\tau_r},
\label{propagation_equation}
\end{eqnarray}
where $\psi=\psi(\mathbf{x},p,t)$ is the CR density per momentum interval, $Q(\mathbf{x}, p)$ is the source term,
$\dot{p}\equiv \mathrm{d}p/\mathrm{d}t$ is the momentum loss rate,
and the time scales $\tau_f$ and $\tau_r$ characterize fragmentation processes and radioactive decays, respectively.

\begin{table*}
\centering
\setlength\tabcolsep{0.4em}
\caption{Propagation models considered in this paper.}
\label{tab:propmodel}
\begin{tabular}{ccc}
\hline\hline
Model & Propagation parameters & Description  \\
 \hline
DR & {$D_0$, $\delta$, $v_A$, $z_h$} & Diffusion + reacceleration, $\eta=1$, $R_0=4$~GV  \\
DR-2 & {$D_0$, $\delta$, $v_A$, $z_h$} & Diffusion + reacceleration, $\eta=-0.4$, $R_0=4$~GV  \\
DC  & {$D_0$, $\delta$, $dV_c/dz$, $R_0$, $z_h$} & Diffusion + convection, $\eta=1$, $\delta=0$ for $R<R_0$ \\
\hline\hline
\end{tabular}
\end{table*}

The convection velocity $\mathbf{V}_c$ is usually
assumed to linearly depend on the distance away from the Galactic disk and the convection effect can be described by the quantity $\mathrm{d}V_c/\mathrm{d}z$.
The spatial diffusion coefficient $D_{xx}$ can be parameterized as \cite{Maurin:2010zp}
\begin{equation}
D_{xx} = D_0\beta^\eta \left( R/R_0 \right)^{\delta},
\end{equation}
where $R\equiv pc/Ze$ is the rigidity, $\beta$ is the CR particle velocity in units of the light speed $c$, and $D_0$ is a normalization parameter. Although the slop of diffusion coefficient $\delta$ is predicted to be $\delta=1/3$ for a Kolmogorov spectrum of interstellar turbulence, or $\delta=1/2$ for a Kraichnan cascade, it is usually taken as a free parameter when explaining data. The factor of $\beta^\eta$ denotes the effect that the diffusion coefficient could be altered at low velocities due to the turbulence dissipation.

The CR reacceleration process due to collisions with interstellar random weak hydrodynamic waves can be described by the diffusion in momentum space with a coefficient $D_{pp}$, which is related with $D_{xx}$ by~\cite{1994ApJ...431..705S}
\begin{equation}
D_{pp} D_{xx}=\frac{4p^2v^2_{A}}{3\delta(4-\delta^2)(4-\delta)\omega },
\label{reacceleration}
\end{equation}
where $v_{A}$ is the Alfv\'{e}n velocity and $\omega$ is the ratio of the magnetohydrodynamic wave energy density to the magnetic field energy density. Since $D_{pp}$ is proportional to $v^2_A/\omega$, we can set $\omega=1$ and just use $v_A$ to characterize the reacceleration effect. Therefore, the major propagation parameters involve
$D_0$, $\delta$, $R_0$, $\eta$, $v_A$, $\mathrm{d}V_c/\mathrm{d}z$, and $z_h$.

When CRs propagate in the solar system, their spectra with $R\lesssim 20$~GV are significantly affected by the solar wind.
This is the solar modulation effect, which depends on the solar activity and varies with the solar cycle.
It can be described by the force field approximation~\cite{Gleeson:1968zza} with only one parameter, i.e., the solar modulation potential $\phi$.

Due to fragmentation and radioactive decays, secondary CR particles are produced in propagation processes.
As a result, secondary-to-primary ratios and unstable-to-stable secondary ratios, such as B/C and $^{10}\mathrm{Be}/^9\mathrm{Be}$, are sensitive to propagation paramors, but almost independent of primary injection spectra.
Thus the measurements of these ratios are useful to determine the propagation parameters (see e.g. Refs.~\cite{Putze:2010zn,Trotta:2010mx}).

Note that there are degeneracies between the propagation models with reacceleration process and those with convection effects. Current measurements are not sufficient to distinguish these models.
Therefore, we separately consider the DR and DC models when calculating the antiproton flux. The descriptions of these models are given in Table~\ref{tab:propmodel}.
$R_0$ is taken to be 4~GV in the DR and DR-2 models.
In the DC model, however, $R_0$ is set to be a free parameter and $\delta=0$ is imposed for $R<R_0$.

\section{Astrophysical prediction for the $\bar{p}/p$ ratio}

In this section, we calculate the secondary antiproton flux generated by CRs when they propagate in the Galaxy considering the uncertainties
from propagation processes. In the following, we first derive the ranges of propagation parameters.

It is generally assumed that CR particles are accelerated in supernova remnants (SNRs).
Thus we assume the spatial distribution of the CR sources follow the SNR distribution as
\begin{equation}
f(r,z) = \left( \frac{r}{r_\odot} \right)^a \exp\left[ -\frac{b(r-r_\odot)}{r_\odot} \right]\exp\left( -\frac{\left|z\right|}{z_s} \right),
\label{spatial_distribution}
\end{equation}
where the distance from the Sun to the Galactic Center $r_\odot=8.5$~kpc, and the characteristic height of the Galactic disk $z_s\simeq 0.2$~kpc.
We adopt $a=1.25$ and $b=3.56$~\cite{Trotta:2010mx} in the calculation. In the DR-2 and DC models, the injection spectra of primary nuclei are taken to be single power-law functions. In the DR model,
the injection spectra are assumed to be broken power-law functions with respect to the rigidity as
\begin{equation}
   \left\{ \begin{array}{ll}
    q_i \propto\left( \dfrac{R}{R_\mathrm{br}} \right)^{-\nu_1}, & R \le R_\mathrm{br};\\
    q_i \propto\left( \dfrac{R}{R_\mathrm{br}} \right)^{-\nu_2}, & R > R_\mathrm{br}.
  \end{array}
  \right.
\label{injection_power_law}
\end{equation}
Then the injection source term is $Q_i({\bf x},p)=f(r,z)q_i(p)$.

\begin{figure*}[!htbp]
\centering
\includegraphics[width=.55\textwidth]{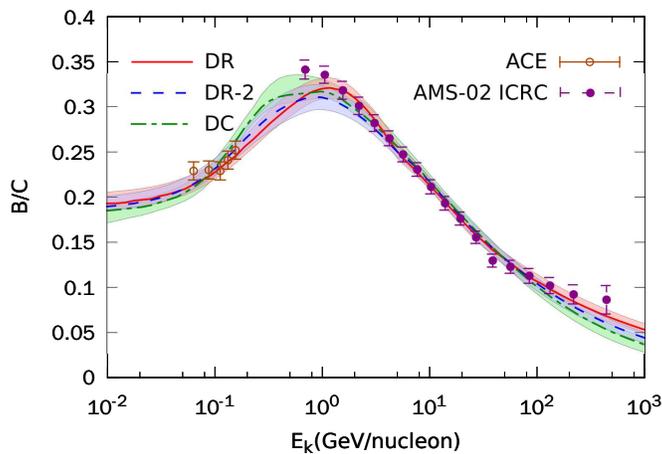}
\caption{Fitting results of the B/C ratio within 95\% confidence ranges in three propagation models, compared with the data from AMS-02~\cite{AMS_ICRC13} and ACE~\cite{Mewaldt:2000zz}.}
\label{fig:BCratio}
\end{figure*}

We use the numerical tool GALPROP~\cite{Moskalenko:1997gh,Strong:1998pw} to solve the propagation equation. The propagation parameters are determined by 
fitting to the nuclear data B/C and $^{10}$Be/$^9$Be.
In order to improve the fitting efficiency, the MCMC method is employed to derive the posterior probability distributions of the propagation parameters.
By fitting the B/C ratio data from AMS-02~\cite{AMS_ICRC13} and ACE~\cite{Mewaldt:2000zz}, and the $^{10}\mathrm{Be}/^9\mathrm{Be}$ ratio data from ACE~\cite{2001ApJ...563..768Y}, Balloon~\cite{1977ApJ...212..262H,1978ApJ...226..355B,1979ICRC....1..389W}, IMP7\&8~\cite{IMP7_8}, ISEE3-HKH~\cite{ISEE3-HKH}, ISOMAX~\cite{Hams:2004rz}, Ulysses-HET~\cite{1998ApJ...501L..59C}, and Voyager~\cite{Voyager}, we obtain the mean values and $1\sigma$ errors of the propagation parameters for the three models, as shown in Table~\ref{tab:BtoC_fit}. We also show the fitting results of B/C ratio within 95\% confidence ranges compared with the experimental data in Fig.~\ref{fig:BCratio}.

\begin{table*}
\centering
\setlength\tabcolsep{0.4em}
\caption{Mean values and $1\sigma$ uncertainties of the propagation parameters derived through fitting the data of B/C  and $^{10}\mathrm{Be}/^9\mathrm{Be}$ ratios in three propagation models.}
\label{tab:BtoC_fit}
\begin{tabular}{cccc}
\hline\hline
 & DR & DR-2 & DC \\
\hline
$D_0$ ($10^{28}~\cm^2~\sec^{-1}$) & $6.58\pm 1.27$ & $3.59\pm 0.88$ & $1.95\pm 0.50$ \\
$\delta$ & $0.333\pm 0.011$ & $0.423\pm 0.017$ & $0.510\pm 0.034$ \\
$R_0$ (GV) &  4 & 4 & $4.71\pm 0.80$ \\
$v_A$ ($\km~\sec^{-1}$) & $37.8\pm 2.7$ & $22.6\pm 3.1$ & / \\
$\mathrm{d}V_c/\mathrm{d}z$ ($\km~\sec^{-1}~\kpc^{-1}$) & / & / & $4.2\pm 3.2$ \\
$z_h$ (kpc) & $4.7\pm 1.0$ & $3.5\pm 0.8$ & $2.5\pm 0.7$ \\
$\phi_\mathrm{B/C}$ (MV) & $326\pm 36$ & $334\pm 37$ & $182\pm 25$ \\
\hline\hline
\end{tabular}
\end{table*}

\begin{figure}[!htbp]
\centering
\includegraphics[width=.55\textwidth]{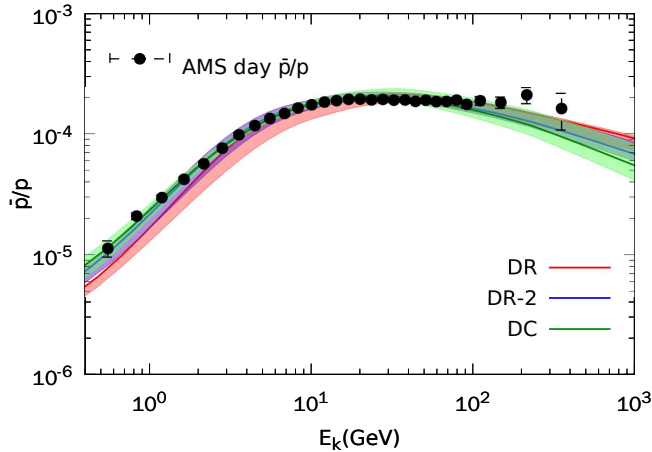}
\caption{Astrophysical predictions of the $\bar{p}/p$ ratio for propagation parameters varying within 95\% confidence ranges in the DR (red band), DR-2 (blue band), and DC (green band) propagation models. Data points denote the AMS-02 $\bar{p}/p$ ratio measurement~\cite{AMSantip}.}
\label{fig:pbar_astro}
\end{figure}

Then the spectrum of secondary antiprotons and $\bar{p}/p$ ratio are calculated taking into account the propagation parameter uncertainties.
Here the proton injection spectrum is determined by fitting to the AMS-02~\cite{AMS_ICRC13} and CREAM~\cite{Ahn:2010gv} data. It should be noted that
the proton spectrum hardening at $\sim 200$ GeV enhances the $\bar{p}/p$ ratio at the high energy end. But even with such proton spectrum hardening
the predicted $\bar{p}/p$ ratio shows a trend that decrease with energy for $E >100$ GeV, which seems not consistent with the AMS-02 data.
The $\bar{p}/p$ ratio predictions for propagation parameters varying within 95\% confidence ranges in the three propagation models are presented in Fig.~\ref{fig:pbar_astro}.
The solar modulation potential $\phi$ are taken as $0.7$ GV for the DR and DR-2 models, and $0.5$ GV for the DC model. 


We notice that the DR model underproduces antiprotons at low energies compared with the AMS-02 data. Actually such an underproduction has been realized for a long time when
studying the PAMELA~\cite{Adriani:2008zq,Adriani:2010rc} and BESS~\cite{Orito:1999re} results \cite{Evoli:2011id,Moskalenko:2001ya,Trotta:2010mx,Hooper:2014ysa}.
In order to solve this problem, the diffusion coefficient at low energies should be modified~\cite{Moskalenko:2001ya}. Other effects, such as the charge-dependent solar modulation and convection, can also be used to avoid this discrepancy~\cite{Hooper:2014ysa,Boudaud:2014qra}. From Fig.~\ref{fig:pbar_astro}, we can see that the DR-2 model with $\eta=-0.4$ and the DC model with a break in the diffusion coefficient can well fit the data up to $\sim 100~\GeV$.
For $E_k \gtrsim 100~\GeV$, however, the predicted $\bar{p}/p$ ratios in all these models tend to decrease, and seem to be hardly reconciled with the AMS-02 data.

\section{Implications for dark matter annihilations}

In this section we study the implications of the AMS-02 $\bar{p}/p$ data on DM annihilation.
As the AMS-02 data at high energies and the propagation model uncertainties are still large, there is no obvious antiproton excess in data. However, the high energy discrepancies between the predicted astrophysical background and the data leave space for contributions from extra antiproton sources. It is usually expected that DM particles in the Galactic halo may annihilate into quarks and gauge bosons, and produce antiprotons after decay and hadronization processes.

For self-conjugate DM particles, the source term of antiprotons from DM annihilations is given by
\begin{equation}
Q(r,p)=\frac{1}{2}\frac{\rho^2(r)}{m_\chi^2}\left<\sigma v\right>
\frac{\mathrm{d}N}{\mathrm{d}E_k},
\label{eq:DM_src}
\end{equation}
where $m_\chi$ is the DM particle mass, and $\left<\sigma v\right>$ is the thermally averaged annihilation cross section.
$\mathrm{d}N/\mathrm{d}E_k$ is the differential number of antiprotons produced per annihilation. We use the results of PPPC 4 DM ID \cite{Cirelli:2010xx} including the electroweak corrections \cite{Ciafaloni:2010ti} to calculate $\mathrm{d}N/\mathrm{d}E_k$.
The DM density distribution $\rho(r)$
is adopted to be the Navarro-Frenk-White (NFW) profile~\cite{Navarro:1996gj}
\begin{equation}
\rho(r)=\frac{\rho_s}{(r/r_s)(1+r/r_s)^2}.
\label{eq:NFW_profile}
\end{equation}
where $r_s=20$~kpc, and $\rho_s$ is derived from the condition that the local DM density at 8.5~kpc from the Galactic Center is 0.3~GeV~cm$^{-3}$.

When studying the DM implications we take the best fit $\bar{p}/p$ background from the bands in Fig. \ref{fig:pbar_astro} and
then add the contributions from DM annihilations to the $\bar{p}/p$ ratio.
Since the $\bar{p}/p$ ratio prediction in the DR model seems inconsistent to the data, we only consider the DR-2 and DC models below.
In order to give the best
fit to the AMS-02 data we introduce a free factor $c_{\bar{p}}$ to normalize the flux of secondary antiprotons. 
This factor represents possible uncertainties in astrophysical antiproton production processes, such as uncertainties from hadronic interactions (see e.g. Refs.~\cite{diMauro:2014zea,Kappl:2014hha}), ISM density distributions, and nuclear enhancement factors from heavy elements. The solar modulation
potential $\phi_{\bar{p}}$ is another free parameter to give best fit to data.

\begin{figure*}[!htbp]
\centering
\includegraphics[width=0.48\textwidth]{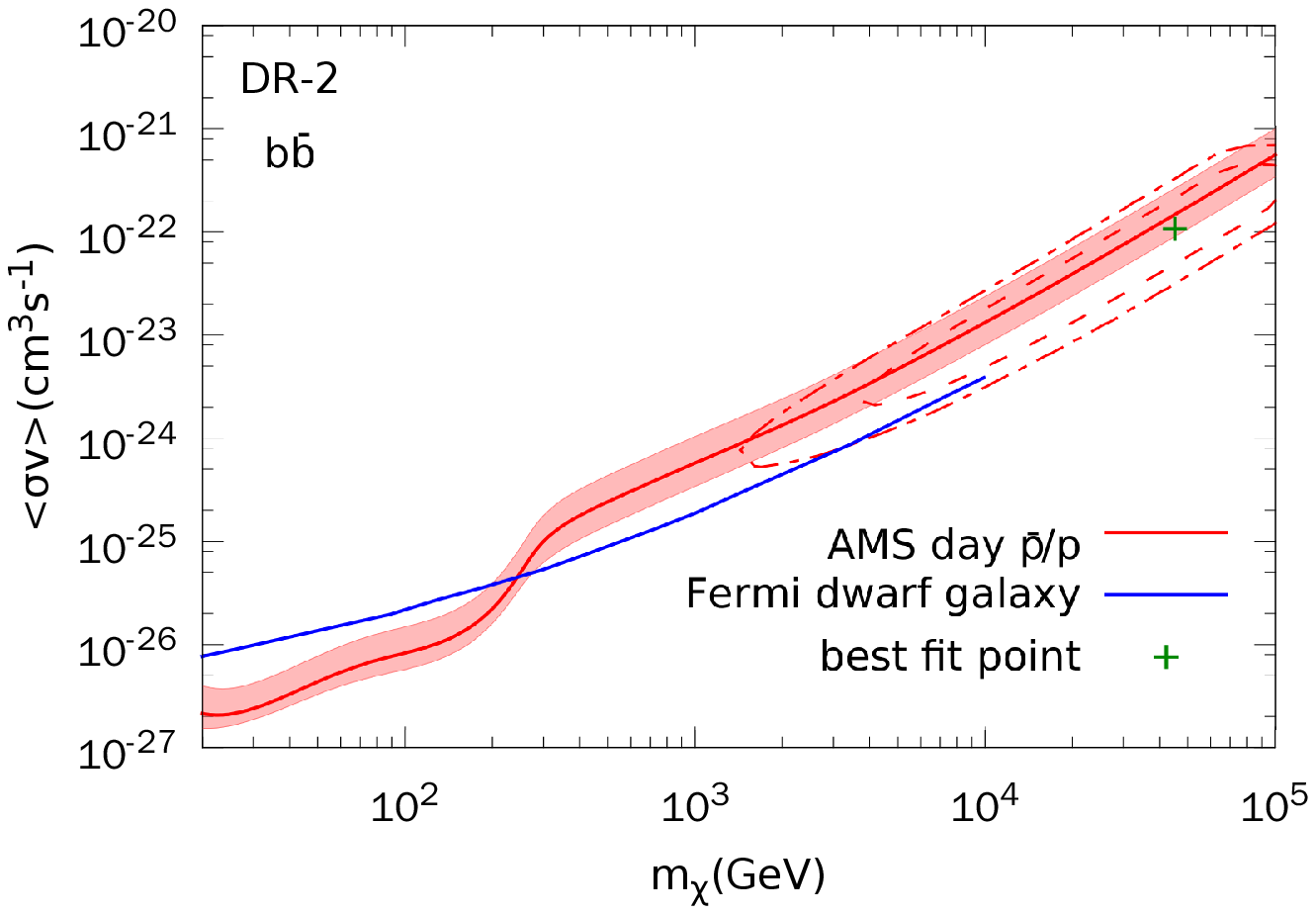}
\includegraphics[width=0.48\textwidth]{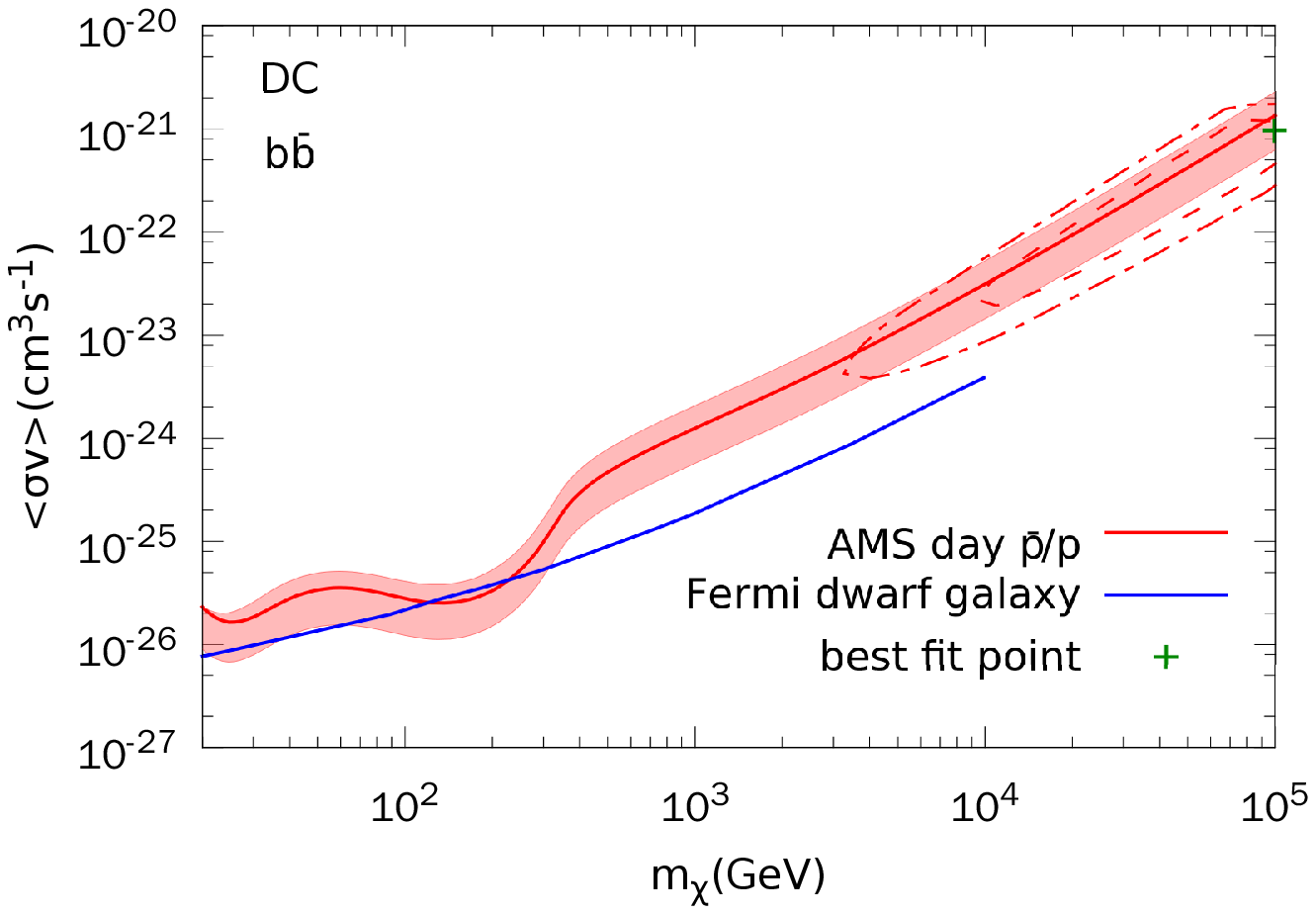} \\
\includegraphics[width=0.48\textwidth]{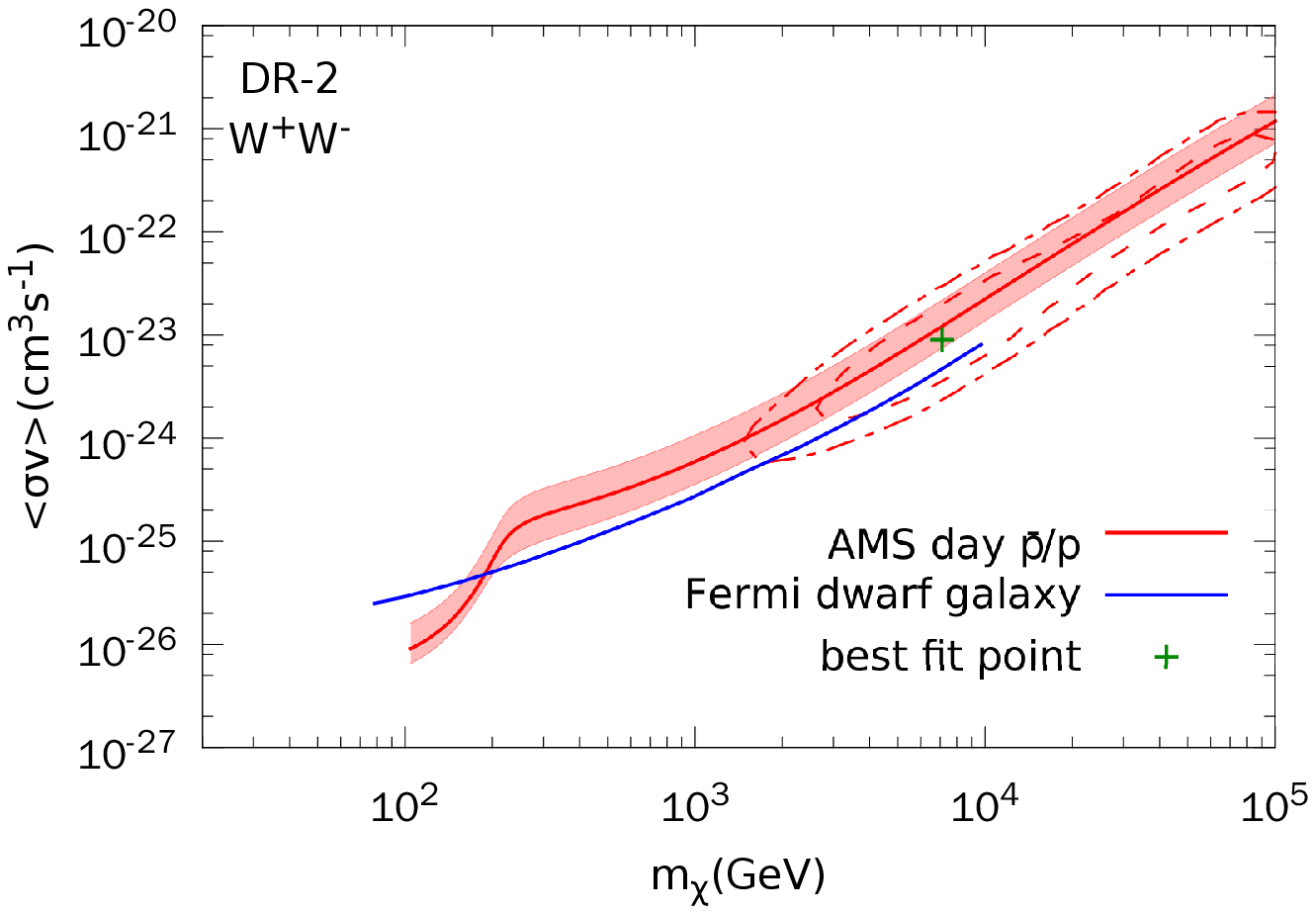}
\includegraphics[width=0.48\textwidth]{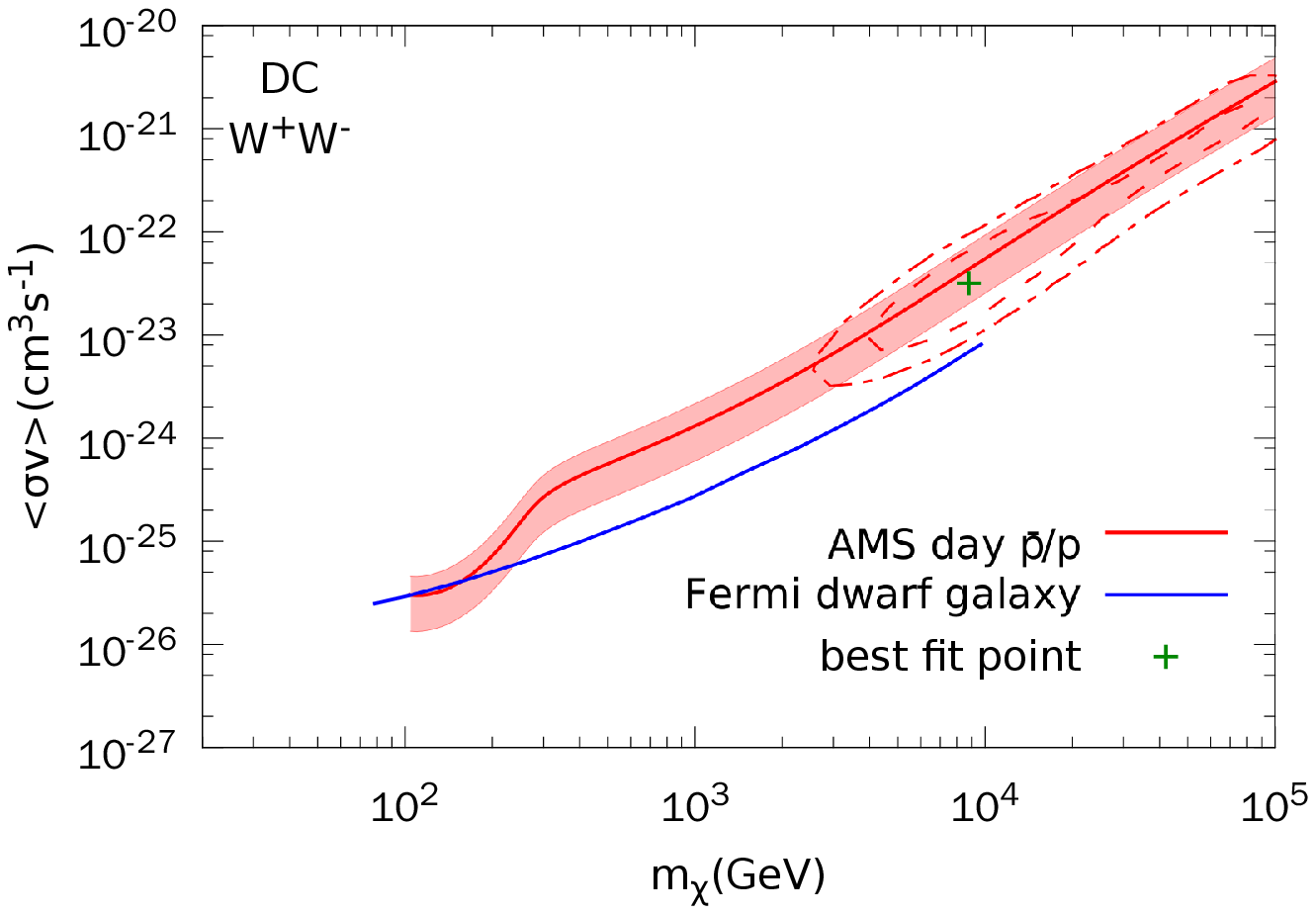}
\caption{The contours denote the 68\% and 95\% confidence regions favored by the AMS-02 $\bar{p}/p$ data in the $m_\chi$-$\left<\sigma v\right>$ plane for the $b\bar{b}$ (upper panels) and $W^+W^-$ (lower panels) annihilation channels. The red band represent the upper limits at 95\% C.L. on the DM annihilation cross section set by the AMS-02 $\bar{p}/p$ data taking the uncertainties from propagation. For a comparison, the upper limits from the Fermi-LAT observations of dwarf galaxies are also shown as blue solid lines. The left and right panels correspond to the results in the DR-2 and DC models, respectively.}
\label{fig:DM_sv}
\end{figure*}

Then we derive the upper limits on the DM annihilation cross section  at 95\% C.L. from the AMS-02 data taking the uncertainties from propagation into account, as shown in Fig. \ref{fig:DM_sv}. Here we consider DM annihilations to $b\bar{b}$ and $W^+W^-$ final states, which are typical cases for antiproton productions. 
The sum of the contributions from background and DM annihilation to the antiproton flux is given by
\begin{equation}
\Phi_{\bar{p}}(m_{\chi}, \left<\sigma v\right>, c_{\bar{p}}, \phi_{\bar{p}})= \Phi_{\rm{bkg}}(c_{\bar{p}}, \phi_{\bar{p}})+\Phi_{\rm{DM}}(m_{\chi}, \left<\sigma v\right>,\phi_{\bar{p}}).
\end{equation}
For each value of $m_{\chi}$, we derive an upper limit on $\left<\sigma v\right>$ by requiring
\begin{equation}
\chi^2(\left<\sigma v\right>)-\chi^2_{\rm{min}}= 3.84,
\end{equation}
where $\chi^2_{\rm{min}}$ corresponds to the best fit.

From Fig.~\ref{fig:DM_sv} we can see that the upper limits in $b\bar{b}$ and $W^+W^-$ channels are very similar due to their analogous initial antiproton spectra. The DR-2 propagation model gives slightly stricter constraints, compared to the DC model. 

\begin{figure*}[!htbp]
\centering
\includegraphics[width=0.48\textwidth]{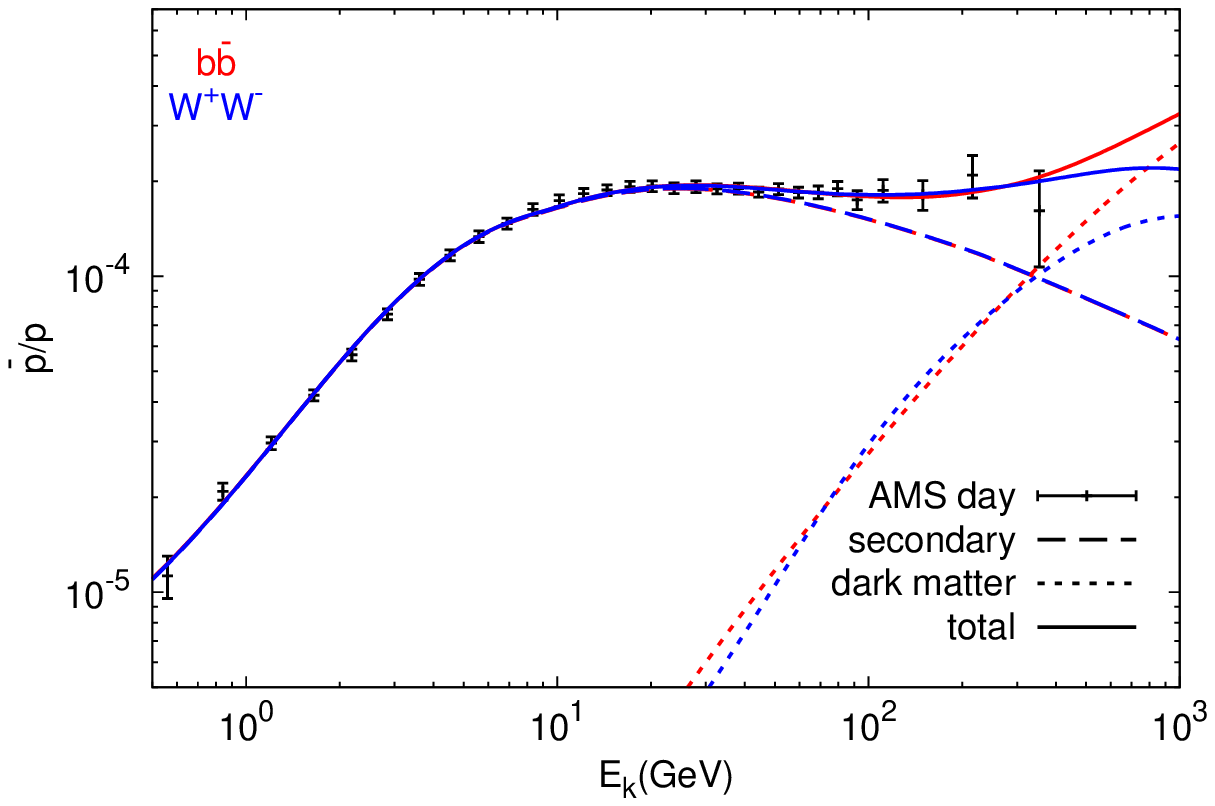}
\includegraphics[width=0.48\textwidth]{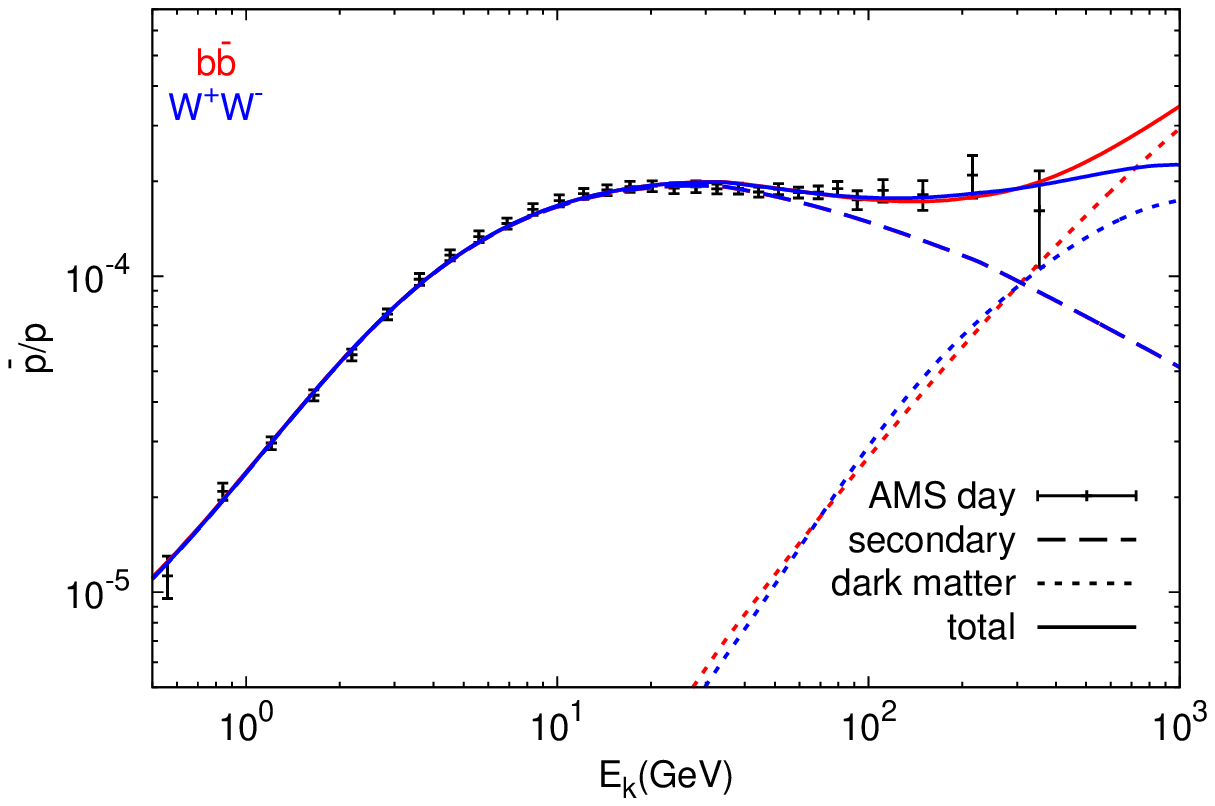}
\caption{$\bar{p}/p$ ratios including the contributions from DM annihilations into $b\bar{b}$ (red lines) and $W^+W^-$ (blue lines) for our best-fit parameters.
Data points denote the AMS-02 $\bar{p}/p$ ratio measurement~\cite{AMSantip}.
The left (right) panel corresponds to the results in the DR-2 (DC) model.}
\label{fig:DM_fit}
\end{figure*}

Since there are discrepancies between the predicted astrophysical backgrounds and the data at high energies, DM contributions to the antiproton flux can significantly improve the fitting to the AMS-02 $\bar{p}/p$ data. We carry out a global fitting by varying \{$c_{\bar{p}}$, $\phi_{\bar{p}}$,  $m_\chi$, $\left<\sigma v\right>$\}, in which $m_\chi$ is set to below $100\,\mathrm{TeV}$, in order to avoid contradicting with the unitarity bound. Then we obtain the best fit values, means values, and $1\sigma$ uncertainties of these parameters, as listed in Table~\ref{tab:DM_parameters}. The $\bar{p}/p$ ratio for the best-fit DM parameters are shown in Fig.~\ref{fig:DM_fit}. 
We can see that $m_\chi\sim \mathcal{O}(10)~\TeV$ and $\left<\sigma v\right>\sim\mathcal{O}(10^{-23})~\cm^3~\sec^{-1}$ could explain the data very well.

\begin{table*}
\centering
\setlength\tabcolsep{0.4em}
\caption{Fitting results for DM annihilations into $b\bar{b}$ and $W^+W^-$ in the DR-2 and DC propagation models.}
\label{tab:DM_parameters}
\begin{tabular}{ccccccc}
\hline\hline
\multicolumn{2}{c}{$b\bar{b}$ channel} & $\log(m_\chi/\GeV)$ & $\log(\left<\sigma v\right>/\cm^3~\sec^{-1})$ & $c_{\bar{p}}$ & $\phi_{\bar{p}}$ (GV) & $\chi^2/\mathrm{dof}$ \\
 \hline
\multirow{2}{*}{DR-2} & best & 4.65 & -21.97 & 0.930 & 0.723 &  11.8/26 \\
 & mean & $4.29\pm 0.45$ & $-22.57\pm 0.72$ & $0.928\pm 0.013$ & $0.724\pm 0.030$ &  \\
\hline
\multirow{2}{*}{DC} & best & 4.99 & -21.02 & 0.945 & 0.696 & 17.6/26  \\
 & mean & $4.48\pm 0.37$ & $-21.89\pm 0.61$ & $0.944\pm 0.012$ & $0.696\pm 0.030$ &  \\
\hline\hline
\multicolumn{2}{c}{$W^+W^-$ channel} & $\log(m_\chi/\GeV)$ & $\log(\left<\sigma v\right>/\cm^3~\sec^{-1})$ & $c_{\bar{p}}$ & $\phi_{\bar{p}}$ (GV) & $\chi^2/\mathrm{dof}$ \\
 \hline
\multirow{2}{*}{DR-2} & best & 3.85 & -23.05 & 0.932 & 0.719 & 10.90/26  \\
 & mean & $4.10\pm 0.46$ & $-22.62\pm 0.81$ & $0.934\pm 0.011$ & $0.717\pm 0.029$ &  \\
\hline
\multirow{2}{*}{DC} & best & 3.94 & -22.50 & 0.947 & 0.691 & 15.7/26 \\
 & mean & $4.15\pm 0.38$ & $-22.15\pm 0.69$ & $0.947\pm 0.011$ & $0.691\pm 0.029$ &  \\
\hline\hline
\end{tabular}
\end{table*}

Fig.~\ref{fig:DM_sv} also shows the favored regions at 68\% and 95\% C.L. in the $m_\chi$-$\left<\sigma v\right>$ plane.
The favored values of $\left<\sigma v\right>$ are highly correlated with the values of $m_\chi$, since the DM source term is proportional to $\left<\sigma v\right>/m_\chi^2$, as indicated in Eq.~\eqref{eq:DM_src}.
Since the DM annihilations into quarks or gauge bosons also produce high energy photons, $\gamma$-ray observations, such as those from Fermi-LAT, would set strong constraints on the DM annihilation cross section. Also shown in Fig.~\ref{fig:DM_sv} are the upper limits from the observations of spheroidal dwarf galaxies with six-year Fermi-LAT gamma-ray data \cite{Ackermann:2015zua}. It can be found that the constraint from AMS-02 is similar to that derived from Fermi-LAT dwarf galaxy observation.

\section{Conclusions}

In this work we investigate the new AMS-02 results of $\bar{p}/p$ ratio. In order to understand the data, the crucial step is to perform a reliable
study on the secondary antiproton flux generated by CR interactions with the ISM. We consider three propagation models, namely
the DR, DR-2 and DC models defined in Tab. \ref{tab:propmodel}. The prorogation parameters are constrained by the B/C data using a MCMC
algorithm. We find that all the models can provide very good fit to the B/C data.

Within $1\sigma$ ranges of the fitted propagation parameters, we give
the predictions of the secondary antiprotons in the three propagation models and compared them with the AMS-02 $\bar{p}/p$ data.
It is found that the DR model underproduces antiprotons at low energies, which has already been noticed for a long time. The DR-2 and DC
models can give a quite good fit to the AMS-02 data below $\sim 100$~GeV. However, it should be emphasized that the predicted secondary $\bar{p}/p$ ratio
 above $\sim 100$ GeV has a different slop from the AMS-02 data. The data keep flat up to $\sim 450$~GeV, but the predictions of all the three models
decrease with the energy. The key factor is that the diffusion coefficient is in a form of single power law, namely $D_{xx} \sim R^{-\delta}$, which is determined
by the B/C data. If the diffusion coefficient $D_{xx}$ permits a flatter behavior at high energies within the allowance of the B/C data, a flatter behavior of $\bar{p}/p$
could also be expected~\cite{Vladimirov:2011rn}.

If we consider an extra source of antiprotons from the Galactic DM annihilation, the fitting to the AMS-02 data can be significantly improved. The data can be well explained by a DM particle with a mass of $\sim 10\TeV$ and an annihilation cross section of $\sim 10^{-23}~\cm^3~\sec^{-1}$ into the $b\bar{b}$ or $W^+W^-$ final states. 

If we assume that there is no obvious excess in the $\bar{p}/p$ ratio, the AMS-02 data can be used to set upper limits on the DM annihilation cross section into $b\bar{b}$ or $W^+W^-$. We find that the constraints given by the DR-2 model are slightly stricter than those given by the DC model. For small $m_\chi$, the constraints from AMS-02 are even stronger than the constraints from Fermi-LAT.

\begin{acknowledgments}
We appreciate Q. Yuan for helpful discussions.
This work is supported by the National Natural Science Foundation of China under
Grants NO. 11475189, 11475191, 11135009, 11175251, and the 973 Program of China under
Grant No. 2013CB837000 and by the Strategic Priority Research Program
``The Emergence of Cosmological Structures'' of the Chinese
Academy of Sciences, under Grant No. XDB09000000.
\end{acknowledgments}



\begin{thebibliography}{99}


\bibitem{AMSantip}
  AMS-02 Collaboration,
  Talks at the 'AMS Days at CERN', 15-17 April, 2015.

\bibitem{Adriani:2008zq}
  O.~Adriani, G.~C.~Barbarino, G.~A.~Bazilevskaya, R.~Bellotti, M.~Boezio, E.~A.~Bogomolov, L.~Bonechi and M.~Bongi {\it et al.},
  Phys.\ Rev.\ Lett.\  {\bf 102}, 051101 (2009)  [arXiv:0810.4994 [astro-ph]].

\bibitem{Adriani:2010rc}
  O.~Adriani {\it et al.}  [PAMELA Collaboration],
  Phys.\ Rev.\ Lett.\  {\bf 105}, 121101 (2010)  [arXiv:1007.0821 [astro-ph.HE]].

\bibitem{Donato:2008jk}
  F.~Donato, D.~Maurin, P.~Brun, T.~Delahaye and P.~Salati,
  Phys.\ Rev.\ Lett.\  {\bf 102}, 071301 (2009)  [arXiv:0810.5292 [astro-ph]].


\bibitem{Evoli:2011id}
  C.~Evoli, I.~Cholis, D.~Grasso, L.~Maccione and P.~Ullio,
  Phys.\ Rev.\ D {\bf 85}, 123511 (2012)  [arXiv:1108.0664 [astro-ph.HE]].


\bibitem{Cirelli:2013hv}
  M.~Cirelli and G.~Giesen,
  JCAP {\bf 1304}, 015 (2013)  [arXiv:1301.7079 [hep-ph]].

\bibitem{Bringmann:2014lpa}
  T.~Bringmann, M.~Vollmann and C.~Weniger,
  Phys.\ Rev.\ D {\bf 90}, no. 12, 123001 (2014)  [arXiv:1406.6027 [astro-ph.HE]].

\bibitem{Cirelli:2014lwa}
  M.~Cirelli, D.~Gaggero, G.~Giesen, M.~Taoso and A.~Urbano,
  JCAP {\bf 1412}, no. 12, 045 (2014)  [arXiv:1407.2173 [hep-ph]].

\bibitem{Hooper:2014ysa}
  D.~Hooper, T.~Linden and P.~Mertsch,
  JCAP {\bf 1503}, no. 03, 021 (2015)  [arXiv:1410.1527 [astro-ph.HE]].

\bibitem{Jin:2014ica}
  H.~B.~Jin, Y.~L.~Wu and Y.~F.~Zhou,
  arXiv:1410.0171 [hep-ph].

\bibitem{Boudaud:2014qra}
  M.~Boudaud, M.~Cirelli, G.~Giesen and P.~Salati,
  arXiv:1412.5696 [astro-ph.HE].


\bibitem{cosmicraybook}
  T.~K.~Gaisser,
  Cosmic rays and particle physics,
  New York, Cambridge University Press, 1990.

\bibitem{Strong:1998pw}
  A.~W.~Strong and I.~V.~Moskalenko,
  Astrophys.\ J.\  {\bf 509}, 212 (1998)  [astro-ph/9807150].

\bibitem{DiBernardo:2009ku}
  G.~Di Bernardo, C.~Evoli, D.~Gaggero, D.~Grasso and L.~Maccione,
  Astropart.\ Phys.\  {\bf 34}, 274 (2010)  [arXiv:0909.4548 [astro-ph.HE]].

\bibitem{Maurin:2001sj}
  D.~Maurin, F.~Donato, R.~Taillet and P.~Salati,
  Astrophys.\ J.\  {\bf 555}, 585 (2001)  [astro-ph/0101231].

\bibitem{Moskalenko:1997gh}
  I.~V.~Moskalenko and A.~W.~Strong,
  Astrophys.\ J.\  {\bf 493}, 694 (1998)  [astro-ph/9710124].


\bibitem{Lewis:2002ah}
  A.~Lewis and S.~Bridle,
  Phys.\ Rev.\ D {\bf 66}, 103511 (2002)  [astro-ph/0205436].

\bibitem{Liu:2009sq}
  J.~Liu, Q.~Yuan, X.~Bi, H.~Li and X.~Zhang,
  Phys.\ Rev.\ D {\bf 81}, 023516 (2010)  [arXiv:0906.3858 [astro-ph.CO]].

\bibitem{Liu:2011re}
  J.~Liu, Q.~Yuan, X.~J.~Bi, H.~Li and X.~Zhang,
  Phys.\ Rev.\ D {\bf 85} (2012) 043507  [arXiv:1106.3882 [astro-ph.CO]].

\bibitem{Yuan:2013eja}
  Q.~Yuan, X.~J.~Bi, G.~M.~Chen, Y.~Q.~Guo, S.~J.~Lin and X.~Zhang,
  Astropart.\ Phys.\  {\bf 60}, 1 (2015)  [arXiv:1304.1482 [astro-ph.HE]].

\bibitem{Lin:2014vja}
  S.~J.~Lin, Q.~Yuan and X.~J.~Bi,
  Phys.\ Rev.\ D {\bf 91}, no. 6, 063508 (2015)  [arXiv:1409.6248 [astro-ph.HE]].


\bibitem{Giesen:2015ufa}
  G.~Giesen, M.~Boudaud, Y.~Genolini, V.~Poulin, M.~Cirelli, P.~Salati, P.~D.~Serpico and J.~Feng {\it et al.},
  arXiv:1504.04276 [astro-ph.HE].


\bibitem{Jin:2015sqa}
  H.~B.~Jin, Y.~L.~Wu and Y.~F.~Zhou,
  arXiv:1504.04604 [hep-ph].


\bibitem{Ibe:2015tma}
  M.~Ibe, S.~Matsumoto, S.~Shirai and T.~T.~Yanagida,
  arXiv:1504.05554 [hep-ph].

\bibitem{Hamaguchi:2015wga}
  K.~Hamaguchi, T.~Moroi and K.~Nakayama,
  arXiv:1504.05937 [hep-ph].  

\bibitem{Strong:2007nh}
  A.~W.~Strong, I.~V.~Moskalenko and V.~S.~Ptuskin,
  Ann.\ Rev.\ Nucl.\ Part.\ Sci.\  {\bf 57}, 285 (2007)
  [astro-ph/0701517].

\bibitem{Maurin:2010zp}
  D.~Maurin, A.~Putze and L.~Derome,
  Astron.\ Astrophys.\  {\bf 516}, A67 (2010)  [arXiv:1001.0553 [astro-ph.HE]].


\bibitem[Seo \& Ptuskin(1994)]{1994ApJ...431..705S}
E.~S.~Seo and V.~S.~Ptuskin,
Astrophys.\ J.\  {\bf 431}, 705 (1994).

\bibitem{Gleeson:1968zza}
  L.~J.~Gleeson and W.~I.~Axford,
  Astrophys.\ J.\  {\bf 154}, 1011 (1968).


\bibitem{Putze:2010zn}
  A.~Putze, L.~Derome and D.~Maurin,
  Astron.\ Astrophys.\  {\bf 516}, A66 (2010)
  [arXiv:1001.0551 [astro-ph.HE]].


\bibitem{Trotta:2010mx}
  R.~Trotta, G.~Johannesson, I.~V.~Moskalenko, T.~A.~Porter, R.~R.~de Austri and A.~W.~Strong,
  Astrophys.\ J.\  {\bf 729}, 106 (2011)
  [arXiv:1011.0037 [astro-ph.HE]].

\bibitem{AMS_ICRC13}
AMS-02 Collaboration,
International CosmicRay Conference,
Rio de Janeiro, Brazil, 2013,
\url{http://www.ams02.org/2013/07/newresults-
from-ams-presented-at-icrc-2013/}.

\bibitem{Mewaldt:2000zz}
  R.~A.~Mewaldt, M.~Miller, J.~R.~Jokipii, M.~A.~Lee, T.~H.~Zurbuchen and E.~Mobius,
  ``Acceleration and transport of energetic particles observed in the heliosphere. Proceedings, Symposium, ACE 2000, Indian Wells, USA, January 5-8, 2000,''
  vol. 528, pp. 421-424.

\bibitem[Yanasak et al.(2001)]{2001ApJ...563..768Y}
N.~E.~Yanasak, M.~E.~Wiedenbeck and R.~A.~Mewaldt {\it et al.},
Astrophys.\ J.\  {\bf 563}, 768 (2001).

\bibitem[Hagen et al.(1977)]{1977ApJ...212..262H}
F.~A.~Hagen,  A.~J.~Fisher and J.~F.~Ormes,
Astrophys.\ J.\  {\bf 212}, 262 (1977).

\bibitem[Buffington et al.(1978)]{1978ApJ...226..355B}
A.~Buffington, C.~D.~Orth and T.~S.~Mast,
Astrophys.\ J.\  {\bf 226}, 355 (1978).

\bibitem[Webber \& Kish(1979)]{1979ICRC....1..389W}
W.~R.~Webber and J.~Kish,
International Cosmic Ray Conference, 1979, Vol.~1, p.~389.

\bibitem{IMP7_8}
M.~Garcia-Munoz, J.~A. Simpson and J.~P.~Wefel,
International Cosmic Ray Conference, 1981, Vol.~2, p.~72.

\bibitem{ISEE3-HKH}
J.~A.~Simpson and M.~Garcia-Munoz,
Space Sci.\ Rev.\ 46, 205 (1988).

\bibitem{Hams:2004rz}
  T.~Hams, L.~M.~Barbier, M.~Bremerich, E.~R.~Christian, G.~A.~de Nolfo, S.~Geier, H.~Gobel and S.~K.~Gupta {\it et al.},
  Astrophys.\ J.\  {\bf 611}, 892 (2004).

\bibitem[Connell(1998)]{1998ApJ...501L..59C}
J.~J.~Connell,
Astrophys.\ J.\ Lett., {\bf 501}, L59 (1998).

\bibitem{Voyager}
A.~Lukasiak,
International Cosmic Ray Conference, 1999,
Vol.~3, p.~41.

\bibitem{Ahn:2010gv}
  H.~S.~Ahn, P.~Allison, M.~G.~Bagliesi, J.~J.~Beatty, G.~Bigongiari, J.~T.~Childers, N.~B.~Conklin and S.~Coutu {\it et al.},
  Astrophys.\ J.\  {\bf 714}, L89 (2010)
  [arXiv:1004.1123 [astro-ph.HE]].

\bibitem{Orito:1999re}
  S.~Orito {\it et al.}  [BESS Collaboration],
  Phys.\ Rev.\ Lett.\  {\bf 84}, 1078 (2000)  [astro-ph/9906426].

\bibitem{Moskalenko:2001ya}
  I.~V.~Moskalenko, A.~W.~Strong, J.~F.~Ormes and M.~S.~Potgieter,
  Astrophys.\ J.\  {\bf 565}, 280 (2002)  [astro-ph/0106567].

\bibitem{Cirelli:2010xx}
  M.~Cirelli, G.~Corcella, A.~Hektor, G.~Hutsi, M.~Kadastik, P.~Panci, M.~Raidal and F.~Sala {\it et al.},
   JCAP {\bf 1103}, 051 (2011)
  [JCAP {\bf 1210}, E01 (2012)]  [arXiv:1012.4515 [hep-ph]].  
  
  
\bibitem{Ciafaloni:2010ti}
  P.~Ciafaloni, D.~Comelli, A.~Riotto, F.~Sala, A.~Strumia and A.~Urbano,
   JCAP {\bf 1103}, 019 (2011)  [arXiv:1009.0224 [hep-ph]].
  

\bibitem{Navarro:1996gj}
  J.~F.~Navarro, C.~S.~Frenk and S.~D.~M.~White,
  Astrophys.\ J.\  {\bf 490}, 493 (1997)
  [astro-ph/9611107].

\bibitem{diMauro:2014zea}
  M.~di Mauro, F.~Donato, A.~Goudelis and P.~D.~Serpico,
  Phys.\ Rev.\ D {\bf 90}, no. 8, 085017 (2014)  [arXiv:1408.0288 [hep-ph]].

\bibitem{Kappl:2014hha}
  R.~Kappl and M.~W.~Winkler,
  JCAP {\bf 1409}, 051 (2014)  [arXiv:1408.0299 [hep-ph]].


\bibitem{Ackermann:2015zua}
  M.~Ackermann {\it et al.}  [Fermi-LAT Collaboration],
  arXiv:1503.02641 [astro-ph.HE].

\bibitem{Vladimirov:2011rn} 
  A.~E.~Vladimirov, G.~e.~Johannesson, I.~V.~Moskalenko and T.~A.~Porter,
  Astrophys.\ J.\  {\bf 752}, 68 (2012)
  [arXiv:1108.1023 [astro-ph.HE]].

\end{thebibliography}
\end{document}